\documentclass[letterpaper]{article}

\usepackage{graphicx}
\usepackage{booktabs}
\usepackage{amsmath}
\usepackage{amssymb}
\usepackage{url}
\usepackage[margin=0.75in]{geometry}
\usepackage{caption}
\usepackage{xcolor}  

\usepackage{flafter}
\usepackage{flushend}  
\usepackage{microtype}  
\graphicspath{{./}}

\twocolumn  

\title{\textbf{Manufactured Divisiveness: Decomposing the Hostile Content
of Seven Social Media Influence Operations}}
\author{Emilio Ferrara\\
Thomas Lord Department of Computer Science\\
University of Southern California\\
\texttt{emiliofe@usc.edu}}
\date{}

\begin{document}
\maketitle

\begin{abstract}
State-backed influence operations are routinely measured as high-prevalence
sources of ``hate'' and ``toxicity.'' We argue those rates rest on a
measurement error: the detectors behind them are validated to catch a broader
definition inclusive of hostility or divisiveness aimed at an out-group, and
so over-attribute hate to content better described as partisan or geopolitical
invective. Across 25.08M tweets from seven government-attributed campaigns in
the Twitter Information Operations archive (8{,}275 accounts), we separate
hate from the other forms of divisiveness. We first validate a two-prompt
LLM-based detector, matching human labels at Cohen's $\kappa=0.82$, to
identify the broader hostility; we then develop an auditable rule, agreeing
with an expert at $\kappa=0.52$, to further classify this content (5{,}457
posts) into three sub-categories. About 50.1\% are identity-based attacks on
people, whereas 30.4\% are partisan attacks and 19.5\% invective against
states and their foreign policy. Reporting all of it as hate therefore
overstates hate roughly twofold; only 18.7\% is both identity-based and
dehumanizing or inciting. Six of seven campaigns sort into three regimes that
a single ``hate'' rate flattens, namely identity hate (RU-op and IRA, both
Russia-attributed), geopolitical invective (both Iran operations), and
partisan divisiveness (both Venezuela operations). We call the shared product
\emph{manufactured divisiveness}. The line to separate these constructs itself
remains unsettled: on the hardest cases three independent human experts agree
only moderately (pairwise $\kappa=0.37$--$0.50$), and the best of nineteen LLM
models tops out at $\kappa=0.601$ against the experts' majority. Our findings
can help redefine the study of hate in the context of influence campaigns and
broader online discourse.
\end{abstract}

\section{Introduction}
\label{sec:intro}

State-backed influence operations on social platforms are widely described,
in both policy discourse and the research literature, as sources of
``hate'' and ``toxicity,'' and more broadly of negative and inflammatory
content \cite{stella2018bots}. This framing is increasingly operationalized
directly: an automated detector is applied to operation content, items are
flagged, and the flagged share is reported as a hate or toxicity rate. For
instance, recent work scores tens of millions of posts across many state
actors with a toxicity classifier and reports the toxic share as a headline
measure \cite{shafin2025toxicity,shafin2025language}. These rates are read as
substantive measures of how hateful an operation is, compared across actors as
evidence of platform harm. Their validity
depends on a question seldom stated explicitly: \emph{what construct does the
detector measure?}

A long line of work in hate-speech detection has warned that ``hate'' is
routinely conflated with the broader categories of offensive,
abusive, or divisive language \cite{davidson2017,founta2018,
fortuna2018,vidgen2020,waseem2017typology,banko2020taxonomy}. A detector validated to separate hostile,
divisive out-group targeting from neutral content is a useful instrument,
but not, without further evidence, a hate-speech detector: if the
validated construct is broad, reporting the flagged share as ``hate''
over-attributes hate speech to content that is partisan or geopolitical invective.

We pursue this concern for a corpus of seven government-attributed influence
campaigns (25.08M tweets, 8{,}275 accounts) from the Twitter Information Operations
archive, with a two-stage design that keeps the broad-construct detector and the
typing step distinct and validates each separately. First, a gate detects the
\emph{broad}
construct (hostile or divisive out-group targeting), validated against human
gold at Cohen $\kappa=0.82$,
precision $0.96$ on the original 100-item set. Second, an \emph{auditable
rule} over a frozen, LLM-derived characterization taxonomy types each
positive item as identity-directed hate, partisan divisiveness, or
geopolitical invective, and flags a dehumanizing/inciting severity tier. The
rule's typing agrees with an independent expert (110 items) at $\kappa=0.52$
(moderate), within the band at which two independent human coders agree on the
contested boundary ($\kappa=0.44$). The typing is a rule over a taxonomy, not a
human judgment that an item ``is hate.''

Typing the positives shows compositional heterogeneity: identity hate is only
half, the defensible hate-speech share is just 18.7\%, and the mix
differs systematically across operations scored comparably under the broad gate
(six of seven sort into three construct regimes; \S\ref{sec:find-mix}). We term
the shared product \emph{manufactured divisiveness}, with identity hate the
narrower construct nested within it and the
dehumanizing/inciting core narrower still.

\paragraph{Scope.}
All findings concern these seven specific, \emph{already-detected and
attributed} campaigns and are associational throughout: we report
co-occurrence, divergence, and concentration, not causation. The absence of a
construct here is not evidence of its absence in any other operation.

\paragraph{Contributions.}
\begin{itemize}
  \item \textbf{(C1) On these seven operations, the validated ``hate'' gate
  measures broad divisiveness.} A gate validated against human gold as a
  detector of hostile/divisive out-group targeting is not a hate-speech
  detector: under our stated typing criteria, only $\sim$19\% of its positives
  are identity-directed and dehumanizing/inciting. On a boundary-enriched gold,
  three independent experts agree only moderately (pairwise
  $\kappa=0.37$--$0.50$), each expert is best matched by a \emph{different}
  model, and no model exceeds $\kappa=0.601$ against the experts' majority: the
  divisive/hate line has no annotator-stable reading a broad flag could be
  assumed to capture (Section~\ref{sec:find-validation}).
  \item \textbf{(C2) Positive composition varies by operation, sorting into
  three construct regimes.} Operations
  scored equally ``hateful'' under the broad gate diverge in positive
  composition along construct type: RU-op and IRA are majority identity hate
  (68\% and 61\%), both Iran-attributed operations are majority geopolitical
  invective (IR-op-A 64\%, IR-op-B 51\%), and both Venezuela-attributed
  operations are majority partisan divisiveness (94\% and 75\%, predominantly
  non-hate under the typing rule). This sorting holds for six of seven; the
  seventh (BD-op) splits evenly and is the honest exception. We report
  these as descriptive compositions of the positive set, not as a tested
  discriminative classifier (Section~\ref{sec:find-mix}).
  \item \textbf{(C3) Content-layer divergence attenuates as the construct is narrowed.}
  The pattern ``content (target\slash narrative) diverges, form
  (intensity\slash dehumanization) converges'' holds, but part of the original
  broad-construct contrast came from the identity-versus-political gap itself: as
  the construct is narrowed, content-layer effect sizes fall at each step while
  form-layer associations stay small. We read this as
  an effect-size trend, not a strict law, given the multi-label dimensions and
  small identity-subset samples (Section~\ref{sec:find-layer}).
  \item \textbf{(C4) Concentration is construct-specific.} Accounts with high
  broad-construct concentration do not all stay concentrated on identity hate: one
  operation's concentration falls once attention is restricted to the identity
  core (Section~\ref{sec:find-conc}).
  \item \textbf{(C5) The RU-op prevalence outlier persists at the defensible core.}
  At the dehumanizing-identity-hate core, only RU-op exceeds a non-trivial
  prevalence floor; the RU-op elevation persists under the narrowed construct
  and becomes more pronounced (Section~\ref{sec:find-prev}).
\end{itemize}

\section{Related Work}
\label{sec:related}

\paragraph{Influence-operation content.}
Studies of state-backed operations have characterized troll content,
specialization, and cross-platform influence
\cite{linvill2020,zannettou2019,zannettou2019trolls,starbird2019,badawy2018}, and the 2016
Russian interference campaign in particular has been analyzed as a manipulation
trace \cite{badawy2018,freelon2022blacktrolls}. The partisan audiences these
campaigns court are themselves measurable at scale: political leaning can be
inferred from profile language and retweet-network structure
\cite{jiang2023retweetbert}, which grounds the audience-tuning claims we make
below. More recent comparative analyses of the same Twitter
Information Operations archive characterize many state operations jointly
(by behavioral fingerprint, dissemination, and cross-operation coordination
structure) \cite{saeed2024unraveling,wang2023interstate} rather than by the
type of hostile content they produce. Coordinated inauthentic behavior, large-scale platform manipulation, and the
broader bot ecosystem provide the structural backdrop
\cite{pozzana2020measuring,ng2026bots,ferrara2020characterizing,tardelli2024}.
Much of this literature reports a single hostility or toxicity rate rather
than decomposing the construct; our contribution is to type the construct
before quantifying it.

\paragraph{Hate, offensive, and divisive language.}
The detection literature has repeatedly documented the conflation of hate with
offensive or abusive language \cite{davidson2017,founta2018,fortuna2020toxic}, proposed
typologies that separate abuse along target and form axes
\cite{waseem2017typology}, surveyed the
fragmentation of definitions \cite{fortuna2018}, and traced how annotation
choices propagate into models \cite{vidgen2020,waseem2016,sap2019risk,sap2022attitudes}. Implicit and
boundary cases are especially hard \cite{elsherief2021}. A parallel line in
political communication distinguishes political incivility from group-targeting
intolerance \cite{rossini2022incivility}, and shows that divisive constructs
such as fear speech can target groups while evading toxicity detectors
\cite{saha2021fearspeech}, precisely the heterogeneity a single broad flag
collapses. The instability extends \emph{within} annotators: on RLHF preference
data, the same annotator often rates semantically equivalent harm prompts
inconsistently, and filtering inconsistent annotators flips the majority harm
label on 18.6\% of prompts \cite{ghafouri2026rlhf}.
That three expert annotators agree pairwise at only $\kappa=0.37$--$0.50$ on a
boundary-enriched gold, while model--expert agreement ranges from near zero to
substantial depending on which expert is asked, is consistent with this body of work and with
a parallel cross-lingual audit reporting an LLM-annotator ceiling at
$\kappa\approx0.42$ on an independent corpus~\cite{ferrara2026crosslingual}: the
divisive/hate boundary is genuinely contested, a property of the construct
rather than of our operation set.

\paragraph{LLMs as annotators.}
Large language models have been shown to be competitive with crowd workers for
text annotation \cite{gilardi2023} and useful for computational social science
\cite{ziems2024}, though their annotation reliability is task-dependent and
benefits from task-specific human validation and careful rule design, especially
for content-moderation judgments \cite{pangakis2025humans,kumar2024moderation,plank2022variation,uma2021disagreement,davani2022disagreements}.
We use an LLM gate and characterization, but, consistent
with the validity concerns above \cite{jacobs2021measurement}, we validate the broad gate against human
gold and treat the typing as a transparent rule, not a model verdict.

\paragraph{Moral and affective framing.}
Moral-foundations theory \cite{graham2009}, work on moralized-emotional
diffusion \cite{brady2017}, and evidence that out-group animosity is associated with elevated
engagement \cite{rathje2021outgroup} motivate our affect cross-validation, in
which LLM affect labels are checked against independent moral and emotion
lexica. The distinction we draw between identity-directed hate and partisan
divisiveness parallels political-science work treating partisan animosity as a
distinct, identity-rooted form of out-group hostility, separable from
group-based hate \cite{iyengar2019affective,finkel2020sectarianism}.

\paragraph{Companion work.}
A sibling study examines cross-\emph{lingual} hate in organic platform content
with an actor-less, layered cultural-contingency framing~\cite{ferrara2026crosslingual};
coordination structure
in the same operations is studied separately~\cite{ferrara2026fivemyths}. The present paper is the
cross-\emph{operation} contribution focused on measurement validity: typing the
construct before counting it.

\section{Data}
\label{sec:data}

We analyze seven government-attributed campaigns released through the Twitter
Information Operations archive~\cite{seckin2025}, totaling 25.08M tweets across
8{,}275 accounts. We refer to operations by pseudonymous labels with short
descriptors only; we never name a specific government as the proven actor
beyond ``state-attributed campaigns released through the Twitter Information
Operations archive.'' The descriptors carry only the archive's own coarse
attribution and serve to keep the analysis at the level of measured content
rather than asserting attribution claims of our own.
Table~\ref{tab:data} summarizes per-operation volume and dominant script.

\begin{table*}[t]
\centering
\small
\begin{tabular}{lrrl}
\toprule
Operation (descriptor) & Tweets & Accounts & Dominant script(s) \\
\midrule
RU-op (anti-Muslim/US-wedge)   &    765{,}246 &   359 & Latin \\
IRA (US-partisan)              &  8{,}768{,}633 & 3{,}479 & Cyrillic, Latin \\
IR-op-A (anti-Israel/Saudi)    &  1{,}122{,}936 &   660 & Arabic/Persian, Devanagari \\
IR-op-B (geopolitical/S-Asia)  &  4{,}447{,}056 & 2{,}201 & Arabic/Persian, Latin \\
VE-op-A (domestic ES)          &  8{,}961{,}788 &   987 & Latin (Spanish) \\
VE-op-B (US-facing EN)         &    984{,}980 &   578 & Latin (English) \\
BD-op (domestic)               &     26{,}214 &    11 & Bengali \\
\midrule
\textbf{Total}                 & \textbf{25{,}076{,}853} & \textbf{8{,}275} & --- \\
\bottomrule
\end{tabular}
\caption{Per-operation volume and dominant script. Tweet totals and
distinct-account counts are computed over the analyzed archive corpus
(\texttt{COUNT(DISTINCT userid)} per operation); rows sum to the totals shown.
Pseudonymous labels are used throughout.}
\label{tab:data}
\end{table*}

\paragraph{Tier-N data handling.}
Tier-N denotes our most restrictive reporting tier: all outputs in this paper are
labels and counts only. We report no raw tweet
text, no example slurs or slur lists, no real account handles, and no full
numeric account identifiers anywhere. Operations and accounts are referred to
by pseudonymous labels.

\section{Methods}
\label{sec:method}

\subsection{Broad-construct gate}
A gate built on an instruction-tuned LLM (Qwen2.5-7B) scores each
(tweet, target) pair for the broad construct of \emph{hostile or divisive
out-group targeting}. An item counts as positive only under a two-prompt
consensus: both a permissively worded prompt (CLEAN) and a stricter one
(STRICT2) must flag it (CLEAN $\wedge$ STRICT2). The gate
scored 40{,}464 (tweet, target) rows; the consensus-positive census
(\texttt{qcons}$=1$) is 5{,}457 rows, with 2{,}634 single-prompt
disagreements and 32{,}373 consensus-negative. The conservative two-prompt
consensus is the census used throughout. This broad construct is what the
original 100-item human gold validates; the downstream typing rule carries its
own expert validation (Section~\ref{sec:method-typing}).

\subsection{Characterization}
Each positive item is characterized along an 11-dimension schema
(target group, influence-operation narrative, threat frame, moral foundation,
emotion, dehumanization, stance intensity, and others). Of the 5{,}457
characterized items, 850 non-English items were machine-translated to English (NLLB, a multilingual translation model);
the affect cross-validation below runs on the 5{,}332 items carrying usable
English text (native or pivoted), and we test that affect signal is present
pre-translation.

\subsection{Construct typing}
\label{sec:method-typing}
We type each positive item by an \textbf{auditable rule} over the
\emph{frozen} target-plus-narrative taxonomy produced by characterization. The
rule is not a human ``is-it-hate'' judgment and does not re-label the human
gold; it deterministically maps taxonomy values to one of three constructs:
\begin{itemize}
  \item \textbf{identity\_hate}: the target is a protected or ascriptive
  group (religion, ethnicity/race, nationality-as-a-people, immigration,
  gender/sexuality), or the narrative is identity-diagnostic. This is hate
  speech proper.
  \item \textbf{political\_divisive}: the target is a political
  affiliation, party, ideology, or named politician. Inflammatory, but
  \emph{not} hate speech, a distinction that mirrors the incivility-versus-intolerance
  boundary in political communication \cite{rossini2022incivility}. Because partisan
  animosity is itself identity-rooted \cite{mason2018uncivil} and partisan invective
  can carry identity-coded (dogwhistle) content \cite{mendelsohn2023dogwhistles},
  classifying every such item as non-hate makes the 50.1\% identity share a lower
  bound rather than a point estimate.
  \item \textbf{state\_geopolitical}: the attack is on a state, regime, or
  foreign policy. Foreign-policy invective.
\end{itemize}
Orthogonally, a \texttt{hard\_core} severity flag is set when content is
dehumanizing (animalistic or mechanistic) \emph{or} inciting (a call to
exclude or harm). Two target categories that are systematically ambiguous
between a group of people and a polity (ethnic/national out-groups and
immigrants/refugees) are not assigned by target alone but arbitrated by the
narrative; and an item otherwise typed as identity hate is re-typed to
\textbf{state\_geopolitical} when its text names a terrorist organization or a
foreign state (e.g.\ ISIS, the Turkish military) without attacking the religious
group as people, since such content is security or foreign-policy invective
rather than group hate. A provenance audit characterizes the basis of each
assignment: 69.0\% of assignments are target-decisive, 30.9\% are
narrative-diagnostic, and only 7 items (0.1\%) are resolved by a fallback rule;
the text-level override re-types 10.8\% of items.

\paragraph{What the typing does and does not claim.}
The rule's decisive input is the comparatively objective \emph{target-group}
attribute rather than a subjective ``is-it-hate'' verdict. This resolves an
apparent tension with the contested annotator agreement reported below
(Section~\ref{sec:find-validation}): the \emph{is-it-hate} verdict on borderline
content is contested even between expert annotators, which the rule does
not attempt; it partitions instead by \emph{whom} the content targets, a more
reliable axis. The resulting split is best read as the rule's partition
\emph{under these stated definitions}, not a ground-truth item-level measurement.
We validate the rule against
an independent expert typing of 110 items (stratified across the decisive and
narrative-arbitrated cases, blinded to the rule's inputs): it agrees with the
expert at Cohen $\kappa=0.52$ (accuracy 68\%), moderate and above its
pre-refinement value ($\kappa=0.42$), with agreement highest on partisan and
antisemitic targets (per-target accuracy 0.78--0.90) and lowest on the ambiguous
ethnic and immigration targets that motivated narrative arbitration. To
establish that this agreement reflects the construct rather than one coder's
idiosyncrasy, a second independent expert relabeled an 80-item subset enriched
for the contestable boundary cases, blinded identically. The two human coders
agree at Cohen $\kappa=0.44$ (95\% CI $[0.26, 0.60]$), and on this harder subset
the rule agrees with each human at least as strongly as the humans agree with
each other ($\kappa=0.45$ and $0.55$); the rule therefore sits within the
human--human agreement band, so the moderate ceiling is a property of the
divisive/hate boundary rather than of the rule. Where the two experts agree
(47 of 74 items), the rule matches that consensus at $\kappa=0.71$, so the
residual disagreement concentrates on exactly the items the experts themselves
contest.

\subsection{Validation}
\label{sec:method-validation}
The broad gate is validated against a human gold set. On the original
100-item representative gold, the broad gate agrees with the human at Cohen
$\kappa=0.82$ and precision $0.96$. On a boundary-enriched 102-item gold that
over-samples the divisive/hate line, three independent expert annotators each
labeled every item under identical blinding (one, Expert-1, with a written
per-item rationale), and a panel of nineteen models each judge the same blinded
English text; we score Cohen $\kappa$~\cite{krippendorff2004reliability}
against each expert and against the experts' two-of-three majority, and report
human--human reliability as pairwise Cohen $\kappa$ (with 5{,}000-resample
bootstrap intervals) and Fleiss $\kappa$.
The experts coded the \emph{broad} construct, so this
measures agreement on the broad gate, not on the typing rule. The
representative $\kappa=0.82$ and the boundary-enriched agreements are not directly
comparable (different samples and references) and should be read as two
validation regimes, not one instrument deteriorating.

\subsection{Affect-label cross-validation}
We test whether the LLM affect labels track independent dictionary signal by
comparing them, on 5{,}332 English-text confirmed items, against MFD2.0~\cite{frimer2019mfd}
moral-foundations, NRC~\cite{mohammad2013nrc} and LIWC2015~\cite{pennebaker2015liwc} emotion, and
(for Russian originals) rusentilex~\cite{loukachevitch2016rusentilex} lexica, using a one-sided Mann--Whitney $U$
test with rank-biserial effect size and BH-FDR correction.

\section{Findings}
\label{sec:findings}

\subsection{Positive composition}
\label{sec:find-broad}
Of the 5{,}457 positive items, the typing rule assigns 50.1\% to
identity hate, 30.4\% to partisan divisiveness, and 19.5\% to geopolitical
invective (Table~\ref{tab:mix}, ``All''). These are compositions \emph{within}
the gate-positive set (a cue-and-target-selected hostile tail), not corpus-wide
rates; prevalence floors over random strata are reported separately in
Section~\ref{sec:find-prev}. The split is not an artifact of the conservative
two-prompt gate. Re-typing under a permissive CLEAN-only census ($+2{,}140$ items)
or a STRICT2-only census ($+494$ items) leaves identity hate a plurality but never
a majority ($41.5\%$ and $48.8\%$, versus $50.1\%$ under the two-prompt consensus).
Relaxing the gate shifts composition \emph{toward} partisan and geopolitical
content; the conservative gate is therefore the most identity-favorable choice, and the
over-attribution finding is, if anything, stronger under a looser gate. Only 21.0\% (1{,}146 items) carry
the dehumanizing/inciting \texttt{hard\_core} flag, and the narrowest
defensible hate-speech core (identity-directed \emph{and}
\texttt{hard\_core}) is just 18.7\% (1{,}023 items). Under these typing
criteria, treating the broad flag as a ``hate rate'' over-states hate-speech
prevalence for these seven operations by roughly 2.0$\times$: counting \emph{all}
identity-typed content as hate still halves the broad flag ($5{,}457$ vs.\
$2{,}733$ items). This over-statement is
robust to where the typing rule draws the identity boundary. We sweep that
boundary between two extremes: the most identity-favorable choice counts every
ascriptive-group target as identity hate, with no narrative or text-level routing
out of identity ($66.9\%$ identity, $3{,}653$ items); the most
identity-conservative routes the systematically ambiguous ethnic-national and
immigration targets out of identity ($45.1\%$, $2{,}461$ items). Across this range
the identity share moves only within $45.1$--$66.9\%$ and the over-statement only
within $1.5$--$2.2\times$; no defensible boundary brings the broad flag close to a
hate rate. Measured against the
narrowest defensible core (identity \emph{and} dehumanizing/inciting) the gap
widens to 5$\times$ ($5{,}457$ vs.\ $1{,}023$ items), though that figure compares
the broad construct to its strictest sub-core and is best read as an upper
bound.

\begin{figure}[t]
\centering
\includegraphics[width=\columnwidth]{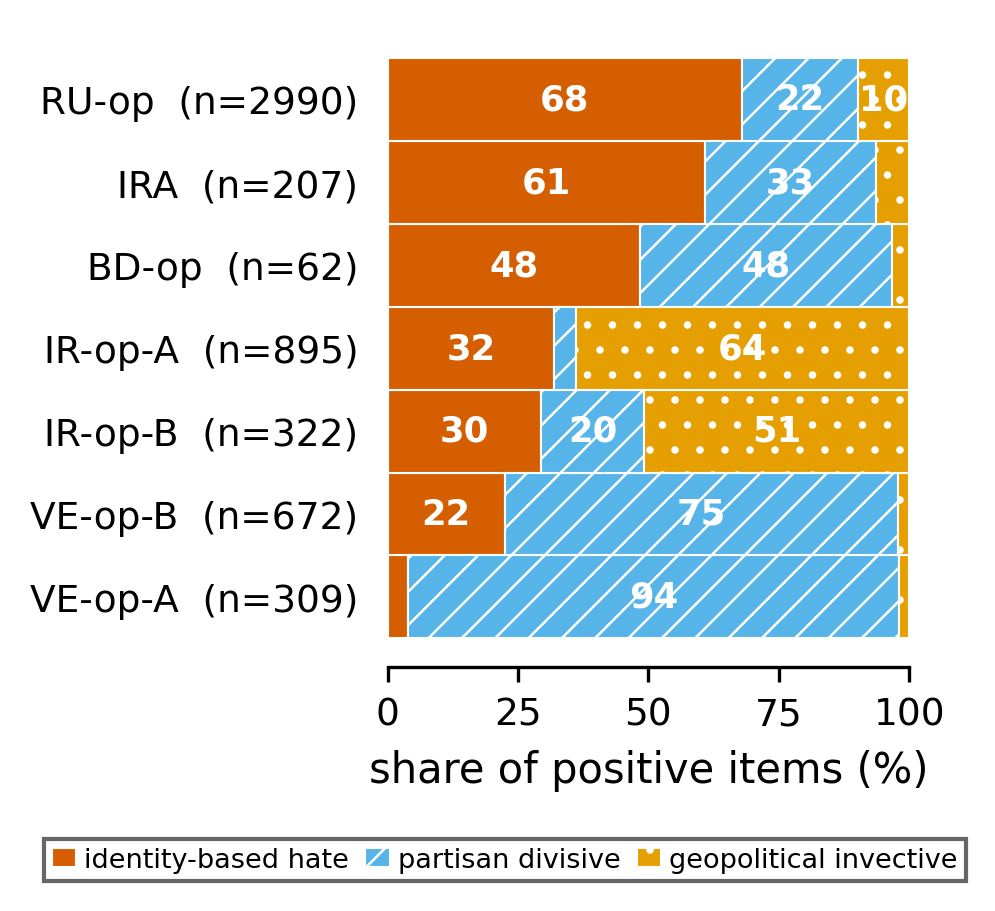}
\caption{Construct composition of positive content, by operation
(row-normalized; $n$ positive items per operation in parentheses). Bars
decompose each operation's hostile content into identity-based hate, partisan
divisiveness, and geopolitical invective. The typing is a rule applied to the
frozen target/narrative taxonomy over items the human-validated broad gate
admitted, not an item-level human classification of hate. The composition
varies by operation: RU-op and IRA are predominantly identity-typed, both
Iran-attributed operations are weighted toward geopolitical invective, and
VE-op-A is almost entirely partisan.}
\label{fig:typing}
\end{figure}

\subsection{Per-operation construct mix}
\label{sec:find-mix}
Operations with comparable positive rates under the broad gate diverge
in composition along construct type, sorting into three regimes
(Fig.~\ref{fig:typing}, Table~\ref{tab:mix}). Two operations are majority
identity hate: RU-op (68.0\%) and IRA (60.9\%). Both Iran-attributed operations
are majority geopolitical invective (IR-op-A 63.8\% geopolitical alongside
32.0\% identity; IR-op-B 50.9\% geopolitical, 29.5\% identity). Both
Venezuela-attributed operations are majority partisan divisiveness: VE-op-A is
94.2\%, with identity hate accounting for only 3.9\% of its positive items, and
VE-op-B is 75.3\%. The seventh, BD-op, splits evenly
between identity hate (48.4\%) and partisan divisiveness (48.4\%). The broad
``hate'' label collapses these compositional differences.

That the regimes align with the polities each operation targets (the
Iran-attributed operations attacking states and foreign policy, the
Venezuela-attributed operations contesting domestic partisan politics, RU-op and
IRA running identity wedges) is unsurprising on its own: different actors have
different adversaries, and the seven operations carry roughly three independent
state attributions, so the regimes rest on a handful of actor families rather than
seven independent draws. The contribution is not that operations differ in target,
which is expected, but that a single broad ``hate'' rate scores all seven alike
while their hostile content belongs to categorically different constructs, a
distinction the scalar rate discards (not a claim that construct type is
determined by actor).

These compositions are row-normalized over each operation's positives, a
selected tail of very different sizes (tens of items for BD-op to thousands),
so the smallest operations' mixes are unstable; we present it as a descriptive
signature, not a tested discriminator (Section~\ref{sec:limitations}).

\begin{table}[t]
\centering
\small
\begin{tabular}{lrrr}
\toprule
Operation & Identity & Partisan & Geopolitical \\
          & hate (\%) & divisive (\%) & (\%) \\
\midrule
RU-op   & 68.0 & 22.2 &  9.8 \\
IRA     & 60.9 & 32.9 &  6.3 \\
IR-op-A & 32.0 &  4.2 & 63.8 \\
IR-op-B & 29.5 & 19.6 & 50.9 \\
VE-op-A &  3.9 & 94.2 &  1.9 \\
VE-op-B & 22.5 & 75.3 &  2.2 \\
BD-op   & 48.4 & 48.4 &  3.2 \\
\midrule
\textbf{All} & \textbf{50.1} & \textbf{30.4} & \textbf{19.5} \\
\bottomrule
\end{tabular}
\caption{Per-operation construct mix (row \%), over the 5{,}457 positive
items.}
\label{tab:mix}
\end{table}

\subsection{The divisive/hate boundary}
\label{sec:find-validation}
Where does ``divisive but not hateful'' end and ``hate'' begin? If that line
were sharp, independent annotators would agree on where it falls. They do not:
three expert annotators of the same borderline items agree pairwise at only
Cohen $\kappa=0.37$--$0.50$ (Fleiss $\kappa=0.45$), unanimous on 63\% of items,
and where they diverge, the models diverge with them.

We test this on two human-labeled gold sets. The \emph{representative} set (100
items) is a random sample of the corpus. The \emph{boundary-enriched} set (102
items) deliberately over-samples items on the divisive/hate line; three experts
independently labeled every item under identical blinding (Expert-1 with a
written per-item rationale), and every annotator is scored against each expert
and against the experts' two-of-three majority. The broad gate agrees with the
representative gold at $\kappa=0.82$ but with the expert majority on the
boundary-enriched set at only $\kappa=0.44$: the instrument did not change; the
second set is stocked with the hardest calls, where a conservative gate and
careful humans part ways.

Would a stronger annotator close that gap? A panel of nineteen models (eleven
open-weight, plus larger open-weight and frontier API models) each redid the
gate task independently on the boundary-enriched gold, judging the same blinded
text (Table~\ref{tab:gold}). Agreement is strikingly heterogeneous, and
\emph{which} model reads the boundary best flips with the expert asked:
Gemini-2.5-pro agrees with Expert-1 at $\kappa=0.705$ but falls to $0.36$ and
$0.48$ against the other two experts, whose best matches are different models
altogether (OLMo-2-7B at $0.55$; OLMo-3.1-32B at $0.53$). Against the
two-of-three expert majority no model exceeds $\kappa=0.601$ (Qwen2.5-14B), with
the deployed gate at $0.44$. There is no capacity ladder: a 7B model (Falcon3)
ties the 32B models, and the newest, most heavily aligned Gemini models agree
\emph{least}, increasingly refusing to call borderline content hostile
\cite{rottger2024xstest} and flagging about 1 in 5 of the items the expert
majority flags (Gemini-3.5-flash $R=0.21$, $\kappa=0.22$).

The one substantial model--expert pairing is thus idiosyncratic rather than a
capacity effect: Gemini-2.5-pro is the most conservative frontier model and
Expert-1 the most conservative coder (30 of 102 positive, against 34 and 42 for
the other two experts), and their $\kappa=0.71$ agreement does not carry over
to either other expert. Expert readings of the boundary differ from one another
(pairwise $\kappa=0.37$--$0.50$) about as much as the better models differ from
any one expert, so no annotator, human or model, supplies a stable reference; a
single ``hate'' rate would inherit whichever reading happened to produce the
gold labels, which is why we report composition rather than a rate. The
difficulty is specific to these boundary items, not the broad construct as a
whole: an eleven-model panel labeling a larger balanced sample agrees at Fleiss
$\kappa=0.47$ on the broad decision, and the strongest challenger reaches
$\kappa=0.72$ on the representative gold (Table~\ref{tab:bench}), so the broad
call itself is reproducible even where the line within it is contested.

\begin{figure}[t]
\centering
\includegraphics[width=\columnwidth]{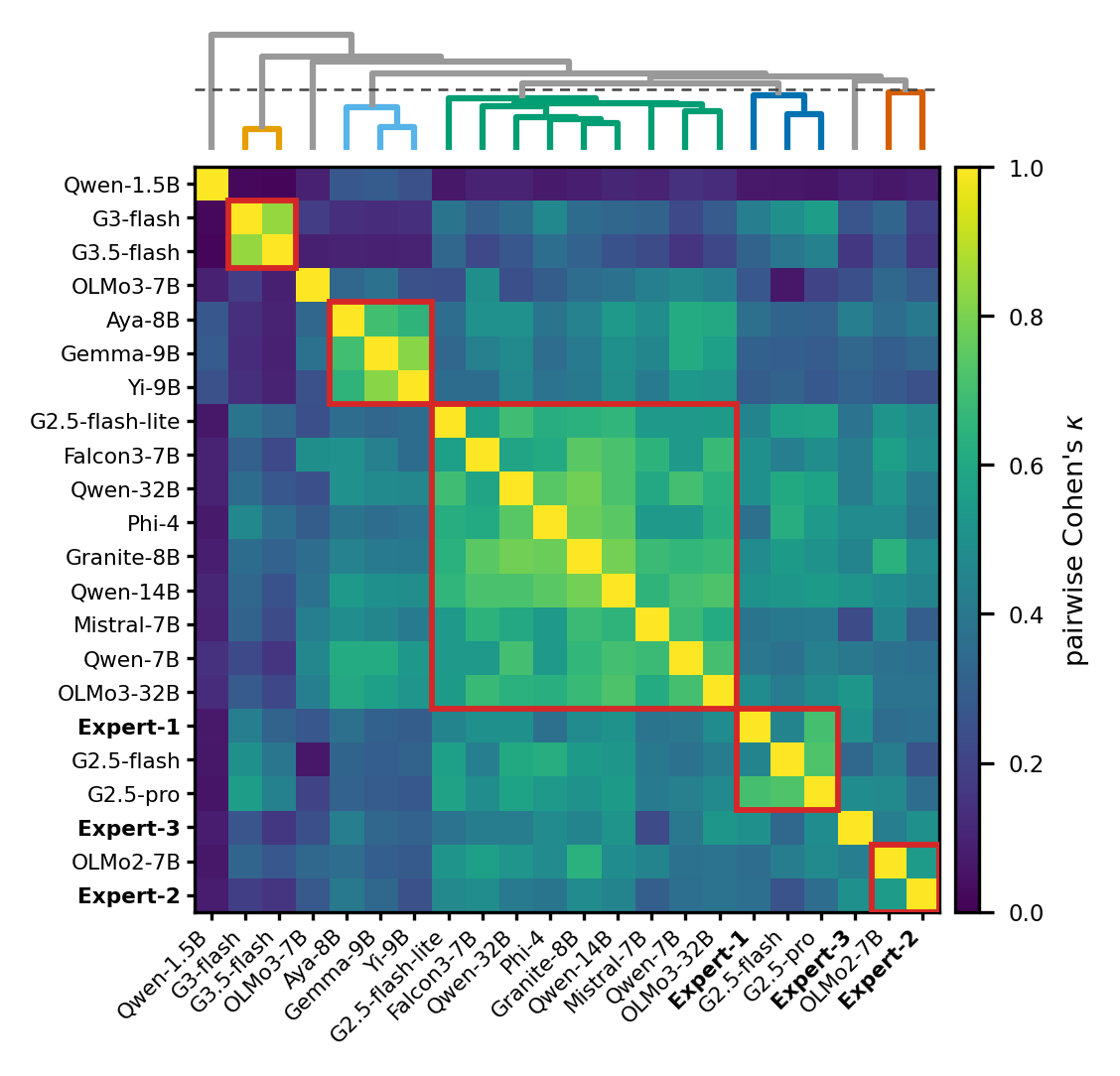}
\caption{Inter-annotator agreement on the broad construct over the
boundary-enriched 102-item gold: pairwise Cohen's $\kappa$ among the nineteen
models and the three human experts, ordered by hierarchical clustering on
$1-\kappa$ (average linkage), so the dendrogram groups annotators that label the
borderline region alike. Red boxes mark the subblocks isolated by a single depth
cut of the dendrogram (dashed line). The three experts (bold) do not group
together: Expert-1 clusters with Gemini-2.5-pro and Gemini-2.5-flash (nearest
neighbor Gemini-2.5-pro, $\kappa=0.71$), Expert-2 clusters with OLMo-2-7B, and
Expert-3 joins no block (nearest neighbors OLMo-3.1-32B and Qwen2.5-14B,
$\kappa=0.52$--$0.53$). Capable mid-sized models form a large
mutually high-agreement block (pairwise $\kappa$ up to $0.84$); the newest Gemini
models and the \{Aya, Gemma, Yi\} group each form their own block; the smallest
model (Qwen2.5-1.5B) sits apart. Darker cells are lower agreement; model sizes
abbreviated, full identifiers in Table~\ref{tab:gold}.}
\label{fig:crossmodel}
\end{figure}

Agreement \emph{among} the annotators tells the same story
(Fig.~\ref{fig:crossmodel}). Clustering the twenty-two annotators (nineteen
models and the three experts) by pairwise Cohen's $\kappa$, the experts do not
group together: Expert-1 clusters with the conservative Gemini-2.5 models (its
nearest neighbor is Gemini-2.5-pro, $\kappa=0.71$), Expert-2 sits beside
OLMo-2-7B, and Expert-3 joins no block at all. Each expert anchors a different
neighborhood, and which annotators fall together is organized by how
conservatively each draws the line, not by scale or recency. The capable
mid-sized models form a mutually high-agreement block (pairwise $\kappa$ up to
$0.84$) that no expert joins.

\begin{table}[t]
\centering
\footnotesize
\setlength{\tabcolsep}{3.4pt}
\begin{tabular}{llrrrr}
\toprule
 & & \multicolumn{3}{c}{$\kappa$ vs expert} & \\
\cmidrule(lr){3-5}
Annotator & Size & E1 & E2 & E3 & Majority \\
\midrule
Qwen2.5-14B & 14B & 0.509 & 0.452 & 0.517 & \textbf{0.601} \\
Granite-3.1-8B & 8B & 0.482 & 0.481 & 0.453 & 0.584 \\
OLMo-2-7B & 7B & 0.356 & \textbf{0.554} & 0.427 & 0.565 \\
\emph{Ensemble (majority)} & -- & 0.502 & 0.427 & 0.503 & 0.543 \\
OLMo-3.1-32B & 32B & 0.485 & 0.379 & \textbf{0.531} & 0.531 \\
Gemini-2.5-pro & API & \textbf{0.705} & 0.356 & 0.484 & 0.531 \\
Phi-4 & 14B & 0.369 & 0.389 & 0.481 & 0.526 \\
Falcon3-7B & 7B & 0.501 & 0.489 & 0.426 & 0.511 \\
Qwen2.5-32B & 32B & 0.501 & 0.407 & 0.426 & 0.511 \\
Gemini-2.5-flash-lite & API & 0.452 & 0.469 & 0.382 & 0.473 \\
Qwen2.5-7B & 7B & 0.397 & 0.366 & 0.400 & 0.440 \\
\emph{Broad gate (2-prompt)} & -- & 0.420 & 0.338 & 0.435 & 0.435 \\
Aya-Expanse-8B & 8B & 0.370 & 0.404 & 0.429 & 0.429 \\
Mistral-7B-v0.3 & 7B & 0.387 & 0.302 & 0.229 & 0.357 \\
Gemini-2.5-flash & API & 0.452 & 0.257 & 0.336 & 0.336 \\
OLMo-3-7B & 7B & 0.270 & 0.279 & 0.245 & 0.331 \\
Gemma-2-9B & 9B & 0.310 & 0.333 & 0.329 & 0.329 \\
Yi-1.5-9B & 9B & 0.296 & 0.247 & 0.316 & 0.316 \\
Gemini-3-flash & API & 0.428 & 0.184 & 0.263 & 0.316 \\
Gemini-3.5-flash & API & 0.318 & 0.151 & 0.162 & 0.216 \\
Qwen2.5-1.5B & 1.5B & 0.068 & 0.078 & 0.082 & 0.082 \\
\bottomrule
\end{tabular}
\caption{Annotator agreement on the broad construct, 102-item boundary-enriched
gold: Cohen's $\kappa$ against each of three independent expert annotators (E1
= P.G., who annotated with per-item rationales; E2, the study's first coder; E3
= D.R.) and against their two-of-three majority, ranked by the majority column.
Rows are the full annotation panel (individual models, their majority
\emph{Ensemble}, and the deployed two-prompt \emph{broad gate}) alongside
larger open-weight and frontier API models run with the byte-identical gate.
Column maxima in bold: each expert is best matched by a \emph{different} model
(Gemini-2.5-pro, OLMo-2-7B, OLMo-3.1-32B), no model exceeds $\kappa=0.601$
against the majority, and the experts themselves agree pairwise at
$\kappa=0.37$--$0.50$ (Fleiss $\kappa=0.45$). Prompt-ablation variants
omitted.}
\label{tab:gold}
\end{table}

\begin{table}[t]
\centering
\small
\begin{tabular}{llrr}
\toprule
Annotator & Size & $\kappa$ & $P$ \\
\midrule
\textbf{Broad gate} \emph{(2-prompt)} & 7B & \textbf{0.820} & 0.960 \\
\midrule
OLMo-3.1-32B & 32B & 0.722 & 0.933 \\
OLMo-3-7B & 7B & 0.623 & 0.886 \\
Gemini-2.5-flash-lite & API & 0.607 & 0.971 \\
Qwen2.5-32B & 32B & 0.605 & 0.923 \\
Gemini-2.5-pro & API & 0.601 & 0.921 \\
Gemini-2.5-flash & API & 0.565 & 0.878 \\
Gemini-3-flash & API & 0.493 & 0.966 \\
Gemini-3.5-flash & API & 0.380 & 0.957 \\
\bottomrule
\end{tabular}
\caption{Best-performing scaled-up and frontier annotators on the
\emph{validated} representative 100-item gold, scored on the broad gate
(two-prompt consensus) and ranked by Cohen's $\kappa$. On this representative
sample models reach up to $\kappa=0.72$, well above their agreement on the
boundary-enriched set: the difficulty is specific to the divisive/hate boundary,
not the broad construct. Greedy decoding, byte-identical gate prompts.}
\label{tab:bench}
\end{table}

\subsection{Validation on operator roles}
\label{sec:find-roles}
The per-operation mix (\S\ref{sec:find-mix}) cannot, on its own, separate
operator intent from language, era, or target availability across seven
heterogeneous campaigns. As an external
check we re-ran the \emph{unchanged} pipeline (the same two-stage gate and
the same construct-typing rule) on a corpus where the operator structure is
independently known: the Clemson/FiveThirtyEight release of Internet Research
Agency tweets \cite{fivethirtyeight2018russiantrolls}, in which every account
carries a hand-assigned role (\emph{Right Troll}, \emph{Left Troll},
\emph{Fearmonger}, \emph{Hashtag Gamer}, \emph{News Feed}, \emph{Commercial})
from \cite{linvill2020}; these released role labels have been treated as ground
truth in subsequent troll-detection work \cite{im2020stillout}. We drew a role-stratified English, non-retweet
candidate sample using the identical cue-and-target selection applied to the
main corpus, gated it, and typed the positive items; roles were withheld
from the model and joined back only afterward. This is concurrent validity
(the broad gate's output checked against an independent, pre-existing labeling of
the same items), within a single actor (the IRA), not independent generalization.

\begin{figure*}[t]
\centering
\includegraphics[width=\textwidth]{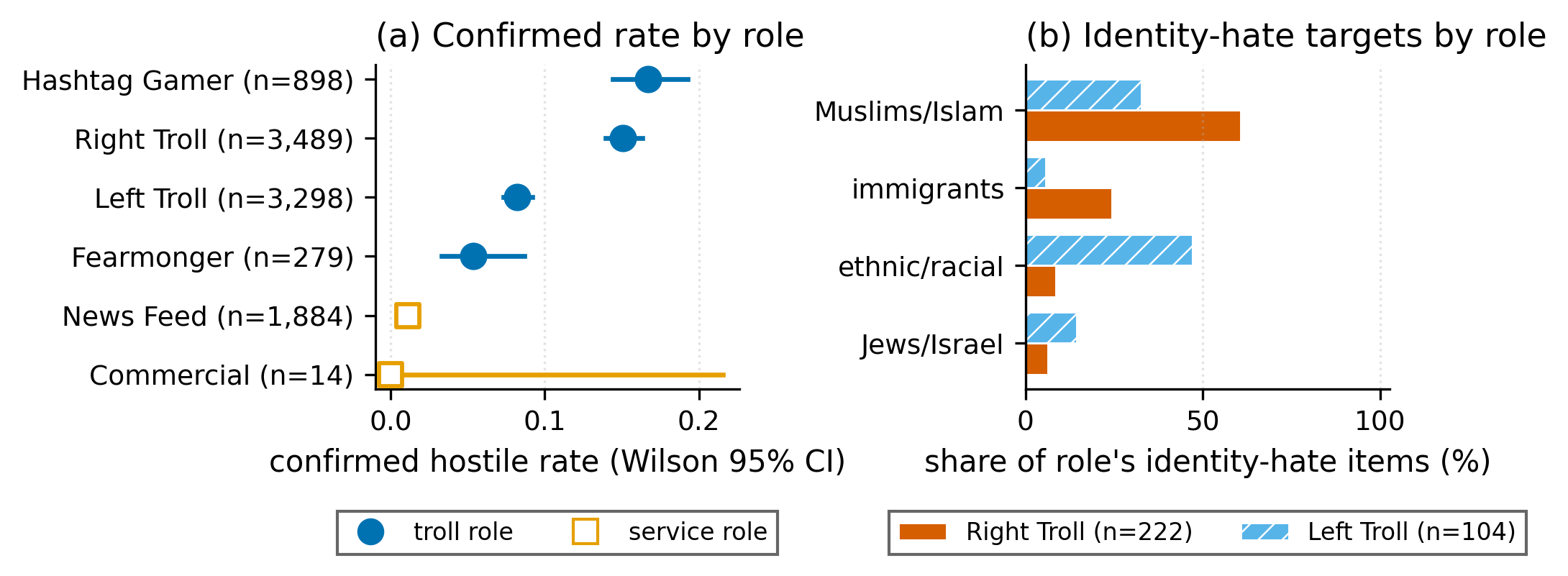}
\caption{Concurrent validation on the Clemson IRA corpus, where operator roles
are independently known \cite{linvill2020,fivethirtyeight2018russiantrolls,arif2018actingthepart};
roles were withheld from the model. \textbf{(a)} The broad gate confirms
hostility on the four troll roles (filled circles) but stays near zero on the
News Feed and Commercial service roles (open squares); bars are Wilson 95\%
intervals, $n$ is candidate items per role. \textbf{(b)} Among identity-hate
items, the right-flank role targets Muslims and immigrants while the left-flank
role centers ethnic/racial out-groups (hatching distinguishes roles in
grayscale). The construct \emph{mix} (identity vs.\ partisan) leans partisan
for all roles (\S\ref{sec:find-roles}); within one operation, roles specialize
by target and intensity, not by construct type. Associational.}
\label{fig:roles}
\end{figure*}

Two signals are associated with the roles (Fig.~\ref{fig:roles}). First, the
broad gate distinguishes the troll roles from the service roles:
confirmed hostility runs 5.4--16.7\% across the four troll roles but falls
to 1.1\% for News Feed and 0.0\% (0/14) for Commercial, the roles whose
function was newswire syndication and advertising, not provocation. Second,
among identity-hate items the roles differ markedly in \emph{whom} they target:
the Right Troll role concentrates on Muslims (60.8\%) and immigrants (24.3\%),
while the Left Troll role centers ethnic/racial out-groups (47.1\%), consistent
with the Islamophobic/anti-immigrant and Black-identity personas documented by
\cite{linvill2020,arif2018actingthepart}.

What does \emph{not} separate the roles is the identity-versus-partisan
construct split itself: among confirmed items the Right Troll role is 42.2\%
identity / 57.4\% partisan and the Left Troll role 38.4\% / 58.7\% (all lean
partisan, as does the Hashtag Gamer role at 32.0\% / 68.0\%), and the
dehumanizing/inciting hard core stays low throughout (5.2--8.7\%). Within one
operation, the roles specialize by \emph{target} and by \emph{intensity}
(operationalized as the confirmation rate), not by construct \emph{type}.
This refines rather than contradicts \S\ref{sec:find-mix}: construct-type
specialization is something
we observe \emph{across} operations, not a property of every division of labor
inside one. (The cue-and-target candidate filter over-selects lexically hostile
content, so these are mixes within a hostile-selected tail, not corpus base
rates.)

\subsection{Prevalence floors by subset}
\label{sec:find-prev}
\begin{figure}[t]
\centering
\includegraphics[width=\columnwidth]{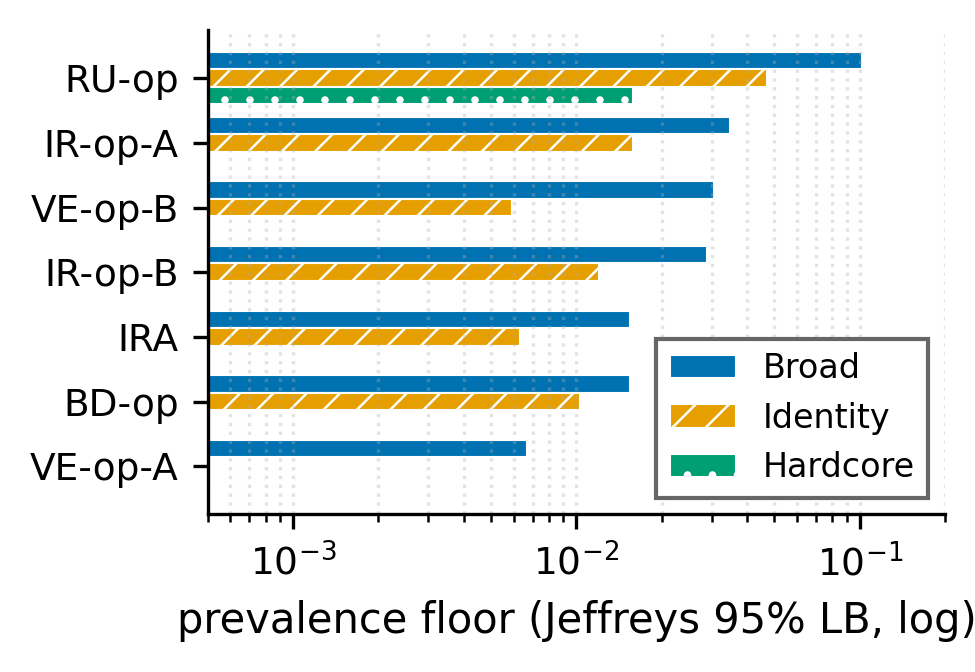}
\caption{Prevalence floors (Jeffreys 95\% lower bound) by construct subset
(broad gate, identity hate, and the dehumanizing hardcore), grouped bars
per operation on a log axis (hatching distinguishes subsets in grayscale).
Floors fall as the construct narrows; RU-op has the highest floor at every
layer, and at the dehumanizing hardcore core only RU-op clears a non-trivial
floor while every other operation sits near zero.}
\label{fig:prev}
\end{figure}

Tweet-level prevalence floors (random stratum, Jeffreys 95\% lower bound,
$n=2{,}000$ per operation) decrease as the construct is narrowed, and the
narrowing is associated with a wider cross-operation contrast (Fig.~\ref{fig:prev}). At the
dehumanizing-identity-hate core, RU-op is the only operation with a
non-trivial floor (0.0158); every other operation is at or near zero, and
VE-op-A's identity floor of 0.0004 is consistent with effectively no identity
hate. The broad-construct RU-op elevation (roughly an order of magnitude above
the lowest operation) thus persists under the narrowed construct and becomes
more distinctly separated from the other operations once the construct is isolated.

\subsection{Contingency under narrowing}
\label{sec:find-layer}
\begin{figure}[t]
\centering
\includegraphics[width=\columnwidth]{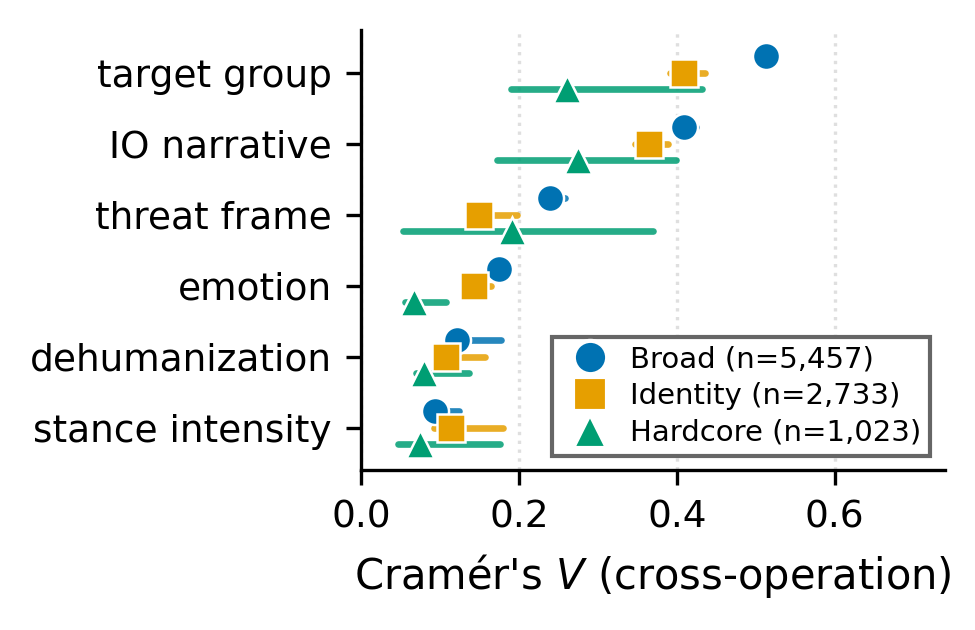}
\caption{Cross-operation Cram\'er's $V$ per dimension, by construct subset
($n$ in legend; rows sorted by the broad-construct value). Each dimension shows three
points (broad construct, circle; identity subset, square; dehumanizing hardcore,
triangle), with bootstrap 95\% confidence intervals (800 resamples) as
whiskers. Content dimensions (target group, IO narrative) diverge substantially
across operations while form dimensions (emotion, dehumanization, stance
intensity) stay low everywhere; the content gap narrows as the construct is
restricted to the identity core, where small $n$ widens the intervals.}
\label{fig:V}
\end{figure}

Across the positive census, operations diverge most on \emph{whom} they
target and \emph{what narrative} they advance, and converge on \emph{how}
hostile they are (stance intensity, dehumanization), reproducing a layered
``content diverges / form converges'' pattern. We measure this with
cross-operation Cram\'er's $V$, an association effect size running from $0$ (no
difference) to $1$ (complete separation): target $0.513$, narrative $0.409$,
intensity $0.093$. Every dimension exceeds its item-level permutation null at
$p<0.001$ ($5{,}000$ permutations); we use this permutation test throughout in
place of an analytic $\chi^2$, whose independence assumption the multi-label
dimensions violate. We compute $V$ two ways. For
dimensions where each item takes exactly one value (target group, stance
intensity, dehumanization), it comes from a standard campaign$\times$category
$\chi^2$ over the $5{,}457$ items. For dimensions where one item can carry
several labels at once (io narrative, threat frame, emotion, moral foundations),
we instead build a campaign$\times$label \emph{incidence} table, counting each
label occurrence separately, so the totals run past the item count (io narrative
$n=6{,}521$, threat frame $5{,}871$, emotion $8{,}052$, moral foundations
$7{,}430$); there $V$ measures how each campaign's mix of labels differs and is
assessed by the same item-level permutation (shuffling the campaign label while
preserving each item's full label set). When the
construct is narrowed, the content-layer divergence
decreases monotonically (Fig.~\ref{fig:V}): target $0.513 \to 0.410 \to
0.260$ and narrative $0.409 \to 0.364 \to 0.275$ across broad, identity, and
hardcore subsets, while intensity stays low and shared. The layered pattern
persists on the dehumanizing-identity-hate core, but part of the original
contrast reflected the identity-versus-political construct gap, which we report
explicitly rather than fold into the layered interpretation.

Temporal variation reinforces this layering. Hostility intensity is temporally
flat (Spearman $\rho\approx0$), consistent with a scripted repertoire that does
not escalate, whereas target composition changes over an operation's active
period: on the identity-hate subset IRA re-aims its targeting substantially
(temporal Cram\'er's $V=0.633$) and IR-op-A moderately ($V=0.410$), consistent
with reports that IRA accounts adapted across the operation's lifespan
\cite{im2020stillout}. Full per-operation trends are in the Supplementary
Information.

\subsection{Account-concentration specificity}
\label{sec:find-conc}
\begin{figure}[t]
\centering
\includegraphics[width=\columnwidth]{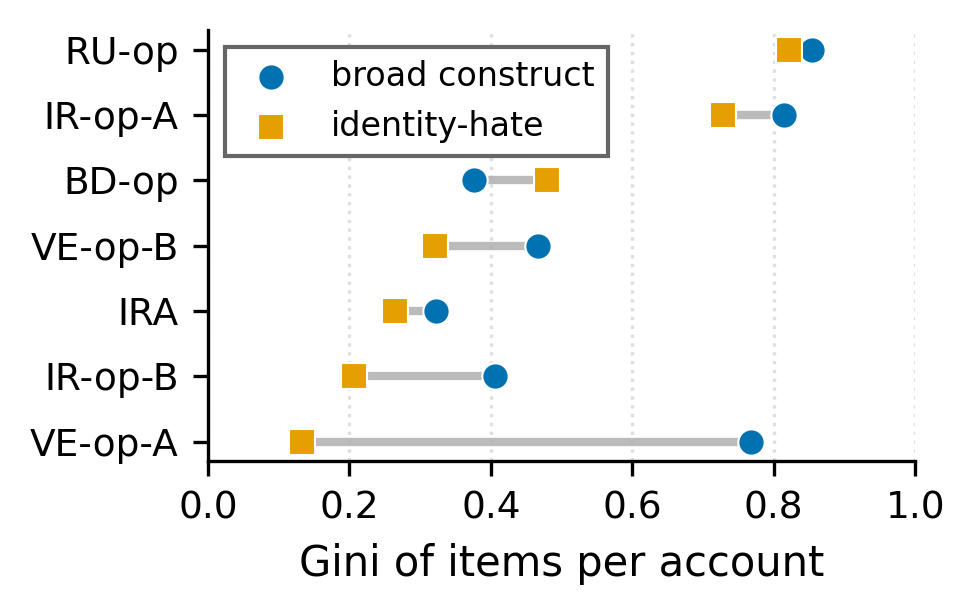}
\caption{Account concentration (Gini of hostile items per posting account),
as a dumbbell from the broad construct to the identity-hate subset. RU-op and
IR-op-A stay highly concentrated at both layers; for VE-op-A concentration
falls ($0.77\!\rightarrow\!0.13$), so its broad-construct concentration
reflects partisan invective rather than identity-hate items.}
\label{fig:conc}
\end{figure}

Account-level concentration of hostile output (the Gini coefficient over
authoring accounts, near $0$ when output is spread evenly across accounts and
near $1$ when it is dominated by a few) is construct-specific
(Fig.~\ref{fig:conc}). Under the broad gate, RU-op
(0.854), IR-op-A (0.814), and VE-op-A (0.768) all exhibit high broad-construct
concentration. But when attention is restricted to identity hate, only RU-op
(0.821) and IR-op-A (0.728) remain highly concentrated on identity hate;
VE-op-A's concentration falls to 0.133, indicating that under the typing
rule its broad-construct concentration is attributable to \emph{partisan-invective} rather than
identity-hate items. So under this rule the broad-construct high-concentration
characterization does not hold for VE-op-A.

\subsection{Affect-label convergent validity}
\label{sec:find-affect}
LLM affect labels are associated with independent lexicon signal to varying degrees (one-sided
Mann--Whitney $U$, with rank-biserial correlation $r_b$ as the effect size, $0$
indicating no separation and $1$ complete separation; BH-FDR). Only care/harm shows a
substantial effect ($r_b=0.520$, on a small positive class of $n_+=111$ of
$5{,}332$); fairness/cheating ($r_b=0.174$) and fear ($r_b=0.105$) are weak
($q<0.001$); anger ($r_b=0.060$, $q=0.01$) and the loyalty/authority labels are
statistically significant but their effect sizes are negligible, with
significance attributable to large $n$. A second, independent emotion lexicon
(LIWC2015~\cite{pennebaker2015liwc}), applied to the same English text,
reproduces the emotion convergence: anger tracks LIWC anger ($r_b=0.096$,
$q=1.4\times10^{-5}$) and fear tracks LIWC anxiety ($r_b=0.134$,
$q<10^{-9}$). Across the confirmed set the LIWC tone score is strongly negative
($23.0$ on a $0$--$100$ scale where below $50$ is negative; negative-emotion
words $4.1\%$ versus positive $1.3\%$), with anger words ($2.3\%$) outweighing
anxiety words ($0.8\%$). Two labels do \emph{not} converge: sanctity/degradation
($r_b=-0.077$, $q=1.0$) and disgust ($r_b=-0.092$, $q=1.0$). On 29
Russian-original confirmed items the rusentilex (a Russian sentiment lexicon) negative-word rate is
$0.026$ (median $0$); we read this small sample as suggestive, not conclusive,
evidence that negativity is present pre-translation. This cross-validation
bears on the moral and emotion fields only; it does not validate the target
taxonomy on which the typing rule operates.

\section{Discussion}
\label{sec:discussion}

\paragraph{Construct typing before counting.}
Methodologically, the construct a detector measures warrants validation, and
content warrants typing before reporting a ``hate rate,'' echoing critiques that
abusive-language constructs are routinely conflated under one label
\cite{banko2020taxonomy}. A gate validated as a broad hostility detector is a
legitimate instrument, but the share it flags is a divisiveness rate, not a hate
rate. For these seven operations, treating the broad flag as ``hate'' over-counts
hate speech by the margins quantified in \S\ref{sec:find-broad}
($2$--$5\times$). This is a
transferable prediction: any broad-hostility detector reported as a hate rate
over-attributes hate by a margin set by how much of the hostile tail is
non-identity invective, and is falsifiable by validating a construct-specific
detector that should show no such gap.

\paragraph{Defining manufactured divisiveness.}
The shared product of these operations is manufactured divisiveness; identity
hate is a narrower construct nested within it, dominant in some operations and
near-absent in others, so that six of seven operations sort into three construct
regimes the broad rate flattens, and the dehumanizing/inciting identity core
is narrower still. Framing it this way connects influence-operation
content to the broader account of affective and sectarian polarization as a
manufactured rather than spontaneous condition \cite{finkel2020sectarianism}.
For operations of this kind (already-detected, attributed campaigns), this favors
reporting composition rather than a single ``hate'' label; we do not extrapolate
beyond the scoped sample.

\paragraph{Re-typing rather than deflation.}
That the broad flag is not a ``hate rate'' must not be read as
``influence-operation hate is overblown.'' We re-\emph{type} hostile content
rather than deflate its prevalence. A real, dehumanizing identity-hate core is
present and, for RU-op, persists under every narrowing of the construct. The
implication for platforms bears first on measurement and reporting: measure the
composition and distinguish the identity-hate core, rather than deprioritizing it on
the basis of the broad label's imprecision. Because the identity boundary that
automated enforcement would key on is the same ambiguous line, we frame this as a
reporting recommendation, with the composition used to triage rather than to set
enforcement thresholds.

\paragraph{Possible sources of compositional divergence.}
The divergence in target and narrative against the convergence in intensity and
dehumanization is consistent with audience-tuned content over a shared production
form. We state this associationally and claim no causal playbook. The IRA role
analysis (\S\ref{sec:find-roles}) fits this reading: where operator structure is
independently known, roles diverge in target and positive rate while sharing a
near-balanced construct mix.

\section{Limitations}
\label{sec:limitations}

The typing is a rule over an LLM-derived taxonomy, not a human adjudication of
whether each item is hate. The broad gate carries human validation
($\kappa=0.82$) and the typing rule is separately validated against an expert
typing of 110 items at moderate agreement ($\kappa=0.52$); the residual
disagreement reflects the ambiguous divisive/hate boundary, on which three
expert annotators themselves agree pairwise at only $\kappa=0.37$--$0.50$, so items
near that boundary are typed by rule, not adjudicated by humans. A second
independent expert relabeled a contested-boundary subset of the typing gold and
agrees with the first only at $\kappa=0.44$, locating this residual disagreement
in the boundary rather than in a single coder (\S\ref{sec:method-typing});
the reported split is the rule's partition under stated definitions, and shifts
under alternative boundary choices within a bounded envelope (identity share
$45.1$--$66.9\%$, over-statement $1.5$--$2.2\times$; Section~\ref{sec:find-broad}).
A further limitation is that the decisive target-group attribute is itself
LLM-derived and is not separately validated at scale, so the partition could
inherit characterization-model bias on that field. The
per-operation construct mix is a descriptive composition of the positive
tail, not a tested discriminative classifier; we fit no null model separating
operator intent from language, era, or target-availability, and the smallest
operations (notably BD-op, 11 accounts) yield unstable per-operation
statistics. Affect cross-validation supports only the moral and emotion fields,
not the target taxonomy that drives the typing; two affect labels (sanctity,
disgust) do not track the reference lexica, and affect on non-English items is
measured after translation, with limited pre-translation
evidence ($n=29$) for Russian. Prevalence figures are floors, not point estimates.
Tier-N reporting (labels and counts only, no released text) means exact item-level
reproduction of the typing requires access to the underlying archive; the typing
rule and the characterization schema are, however, fully specified, so the
partition is auditable in principle by any holder of the source data.
Finally, all findings are scoped to these seven specific, already-attributed
campaigns: the absence of a construct here is not evidence of its absence in
influence operations generally.

\section{Ethics and Data Statement}
\label{sec:ethics}

This work is a secondary analysis of a publicly released, state-attributed
archive and is exempt from human-subjects review; we attribute no actor beyond
the archive's own labeling and report results associationally, scoped to these
seven campaigns. Following the Tier-N protocol (Section~\ref{sec:data}), we
release labels and counts, not tweet text or account identifiers; the full
instrument (verbatim prompts, the typing rule as code, model identifiers, and
seeds) is in the Supplementary Information and reproduces every count.

\paragraph{Acknowledgments.}
Work partly supported by NSF HCC award 2331722. We thank Leonardo Blas, Patrick Gerard, and Daniel Ruiz (USC) for
providing their data annotation expertise.


\newpage
\bibliographystyle{plain}
\bibliography{refs}
\end{document}